\documentclass[10pt,a4paper,twocolumn]{article}
\usepackage{f1000_styles}

\definecolor{paleblue}{HTML}{7892A4}
\definecolor{darkblue}{HTML}{45596E}
\definecolor{midblue}{HTML}{4D637A}

\usepackage{graphicx,wrapfig}
\usepackage{atlast}
\usepackage{enumitem}
\usepackage{hyperref}
\hypersetup{colorlinks,linktocpage,linkcolor={darkblue}, citecolor={darkblue}, urlcolor={darkblue}}

\begin{document}
\pagestyle{fancy}

\title{Atacama Large Aperture Submillimeter Telescope \mbox{(AtLAST)} Science: Probing the Transient and Time-variable Sky}

\author[1]{John Orlowski-Scherer}
\affil[1]{Department of Physics and Astronomy, University of Pennsylvania, 209 South 33rd Street, Philadelphia, PA, 19104, USA}

\author[2]{Thomas J. Maccarone}
\affil[2]{Department of Physics \& Astronomy, Texas Tech University, Box 41051, Lubbock TX, 79409-1051, USA }

\author[3]{Joe Bright}
\affil[3]{Astrophysics, Department of Physics, University of Oxford, Keble Road, Oxford OX1 3RH, UK}
\author[4]{Tomasz Kamiński}
\affil[4]{Nicolaus Copernicus Astronomical Center of the Polish Academy of Sciences, Rabiańska 8, 87100 Toruń, Poland}
\author[5]{Michael Koss}
\affil[5]{Eureka Scientific, 2452 Delmer Street Suite 100, Oakland, CA 94602-3017, USA}
\author[6, 7]{Atul Mohan}
\affil[6]{NASA Goddard Space Flight Center, Greenbelt, MD 20771, USA}
\affil[7]{The Catholic University of America, Washington, DC 20064, USA}
\author[8]{Francisco Miguel Montenegro-Montes}
\affil[8]{Departamento de F{\'i}sica de la Tierra y Astrof{\'i}sica e Instituto de F{\'i}sica de Part{\'i}culas y del Cosmos (IPARCOS). Universidad Complutense de Madrid, Av.\ Complutense, s/n, 28040 Madrid, Spain}
\author[9]{Sigurd Næss}
\affil[9]{Institute of Theoretical Astrophysics, University of Oslo, P.O. Box 1029, Blindern, 0315 Oslo, Norway}
\author[10]{Claudio Ricci}
\affil[10]{Instituto de Estudios Astrofísicos Facultad de Ingeniería y Ciencias Universidad Diego Portales (UDP), Santiago de Chile, Chile}
\author[11]{Paola Severgnini}
\affil[11]{INAF – Osservatorio Astronomico di Brera, Via Brera 28, 20121 Milano, Italy}
\author[12]{Thomas Stanke}
\affil[12]{Max-Planck-Institut f\"ur Extraterrestrische Physik, Gie\ss{}enbachstra\ss{}e 1, D-85748 Garching bei M\"unchen, Germany}
\author[13,14]{Cristian Vignali}
\affil[13]{Dipartimento di Fisica e Astronomia ``Augusto Righi", Universit\`a degli Studi di Bologna, Via Gobetti 93/2, I-40129 Bologna, Italy}
\affil[14]{INAF -- Osservatorio di Astrofisica e Scienza dello Spazio di Bologna, Via Gobetti 93/3, I-40129 Bologna, Italy}
\author[9,15]{Sven Wedemeyer}
\affil[15]{Rosseland Centre for Solar Physics, University of Oslo, Postboks 1029 Blindern, N-0315 Oslo, Norway}

\author[16]{Mark Booth}
\affil[16]{UK Astronomy Technology Centre, Royal Observatory Edinburgh, Blackford Hill, Edinburgh EH9 3HJ, UK}
\author[9]{Claudia Cicone}
\author[17,18,19,20]{Luca Di Mascolo}
\affil[17]{Laboratoire Lagrange, Université Côte d'Azur, Observatoire de la Côte d'Azur, CNRS, Blvd de l'Observatoire, CS 34229, 06304 Nice cedex 4, France}
\affil[18]{Astronomy Unit, Department of Physics, University of Trieste, via Tiepolo 11, Trieste 34131, Italy}
\affil[19]{INAF -- Osservatorio Astronomico di Trieste, via Tiepolo 11, Trieste 34131, Italy}
\affil[20]{IFPU -- Institute for Fundamental Physics of the Universe, Via Beirut 2, 34014 Trieste, Italy}
\author[21]{Doug Johnstone}
\affil[21]{NRC Herzberg Astronomy and Astrophysics, 5071 West Saanich Rd, Victoria, BC, V9E 2E7, Canada}
\author[22]{Tony Mroczkowski}
\affil[22]{European Southern Observatory (ESO), Karl-Schwarzschild-Strasse 2, Garching 85748, Germany}

\author[23]{Martin A. Cordiner}
\affil[23]{Astrochemistry Laboratory, Code 691, NASA Goddard Space Flight Center, Greenbelt, MD 20771, USA.}
\author[12]{Jochen Greiner}
\author[22,24,25]{Evanthia Hatziminaoglou}
\affil[24]{Instituto de Astrof\'{i}sica de Canarias (IAC), E-38205 La Laguna, Tenerife, Spain}
\affil[25]{Universidad de La Laguna, Dpto. Astrof\'{i}sica, E-38206 La Laguna, Tenerife, Spain}
\author[22]{Eelco van Kampen}
\author[16]{Pamela Klaassen}
\author[26,27]{Minju M. Lee}
\affil[26]{Cosmic Dawn Center (DAWN), Denmark}
\affil[27]{DTU-Space, Technical University of Denmark, Elektrovej 327, DK2800 Kgs. Lyngby, Denmark}
\author[12,28]{Daizhong Liu}
\affil[28]{Purple Mountain Observatory, Chinese Academy of Sciences, 10 Yuanhua Road, Nanjing 210023, China}
\author[29,30]{Amélie Saintonge}
\affil[29]{Department of Physics and Astronomy, University College London, Gower Street, London WC1E 6BT, UK}
\affil[30]{Max-Planck-Institut f\"ur Radioastronomie (MPIfR), Auf dem H\"ugel 69, D-53121 Bonn, Germany}
\author[31]{Matthew Smith}
\affil[31]{School of Physics \& Astronomy, Cardiff University, The Parade, Cardiff CF24 3AA, UK}
\author[32]{Alexander E. Thelen}
\affil[32]{Division of Geological and Planetary Sciences, California Institute of Technology, Pasadena, CA 91125, USA.}

\maketitle
\thispagestyle{fancy}

\clearpage
\begin{abstract}

The study of transient and variable events, including novae, active galactic nuclei, and black hole binaries, has historically been a fruitful path for elucidating the evolutionary mechanisms of our universe. The study of such events in the millimeter and submillimeter is, however, still in its infancy. Submillimeter observations probe a variety of materials, such as optically thick dust, which are hard to study in other wavelengths. Submillimeter observations are sensitive to a number of emission mechanisms, from the aforementioned cold dust, to hot free-free emission, and synchrotron emission from energetic particles. Study of these phenomena has been hampered by a lack of prompt, high sensitivity submillimeter follow-up, as well as by a lack of high-sky-coverage submillimeter surveys. In this paper, we describe how the proposed Atacama Large Aperture Submillimeter Telescope (AtLAST) could fill in these gaps in our understanding of the transient universe. We discuss a number of science cases that would benefit from AtLAST observations, and detail how AtLAST is uniquely suited to contributing to them. In particular, AtLAST's large field of view will enable serendipitous detections of transient events, while its anticipated  ability to get on source quickly and observe simultaneously in multiple bands make it also ideally suited for transient follow-up. We make theoretical predictions for the instrumental and observatory properties required to significantly contribute to these science cases, and compare them to the projected AtLAST capabilities. Finally, we consider the unique ways in which transient science cases constrain the observational strategies of AtLAST, and make prescriptions for how AtLAST should observe in order to maximize its transient science output without impinging on other science cases.

\end{abstract}

\section*{\color{OREblue}Keywords}
Time domain --- transient phenomena -- variability -- submillimeter

\clearpage
\pagestyle{fancy}

\section*{Plain language summary}

A wide range of objects and processes observable in the universe vary temporally on scales ranging from minutes to years. 
On timescales of a few hours, cosmic magnetic fields can change their configurations, releasing energy as radiation and particle streams. On time scales of weeks to years, young stars are known to vary in luminosity as they accrete mass. 

The submillimeter radio wavelength regime provides a unique opportunity to observe these phenomena, being sensitive to processes that only emit in that range or are otherwise hidden by dust. So far, however, this field has been limited by a combination of insufficient sky area coverage, poor sensitivity, or poor spatial resolution. Contemporary facilities providing wide field views have small collecting areas, limited sensitivity, and poor spatial resolution. On the other hand, interferometric telescopes such as the Atacama Large Millimeter Array (ALMA) are sensitive and have very fine spatial resolution, but are limited by small fields of view.

The Atacama Large Aperture Submillimeter Telescope (AtLAST) will enable major breakthroughs in the exploration of time-variable phenomena in the submillimeter wavelength regime. Thanks to its 50 m diameter, it will resolve details $\sim 3\times$ smaller than current single-dish telescopes. At the same time, AtLAST will outdo both single-dish telescopes and interferometers in terms of field of view (FoV) by a factor of several thanks to its instantaneous $>1\,\text{deg}^2$ reach.


In this paper we present scientific cases where AtLAST will be transformational in our investigation of the time variable submillimeter sky. We summarize how these observations will help us to better understand fundamental processes in astrophysics, such as the formation of stars, their death, and the relics of their passing. We will also lay out the demands that these observations impose on the design of the telescope and its operations.

\section{Introduction}
The discovery of transient and variable sources in the millimeter and submillimeter wavelength range is in its infancy, with only a few pioneering works \citep{Herzceg+2017, Guns2021, Lee+21, Hervias2023, Hood2023, Li2023} offering a glimpse into this realm. On the other hand, observations of previously known variable objects such as X-ray Binaries, Gamma-Ray Bursts, Supernovae, Tidal Disruption Events, variable Active Galactic Nuclei, and protostars at the short timescales needed to track their  variations (i.e.\ days or shorter) are now a reality with telescopes such as ALMA. So far, submillimeter facilities have mostly been restricted to following up transients discovered at other wavelengths, with relatively few novel detections at these wavelengths \citep{Guns2021, Lee+21, Li2023}. While upcoming observatories such as the Simons Observatory \citep{SO2019} Large Aperture Telescope \citep[SO LAT;][]{Zhu2021} and the CCAT Observatory Fred Young Submillimeter Telescope \citep[CCAT;][]{CCAT2023} promise to systematize the study of the transient submillimeter sky, AtLAST \citep[\url{http://atlast-telescope.org/;}][]{Klaassen2020,Ramasawmy2022,Mroczkowski2023,Mroczkowski2024} will provide a truly revolutionary increase in sensitivity, spatial resolution, and FoV over even next generation experiments. Further, it will be perfectly poised to complement upcoming experiments such as Vera C. Rubin \citep{Hambleton2023}, enabling multi-chroic follow-up and characterization of transient events.

\subsection{Prior/existing instruments and results}

The current state of the art in studying the transient submillimeter and millimeter sky is very limited, in particular when considering blind searches for transient events. Cosmic Microwave Background (CMB) telescopes have so far provided the most complete picture of the transient millimeter sky. The 6-meter Atacama Cosmology Telescope (ACT) has published a blind transient search, detecting 14 events associated with 11 objects, all of which are flaring stars \citep{Li2023}. Further, a targeted search by ACT of known tidal disruption events, supernovae, and other sources yielded only one transient event detected at $5\sigma$, associated with an Active Galactic Nucleus \citep[AGN][]{Hervias2023}. Both of these searches with ACT comprised a decade of data from a state-of-the-art millimeter facility, yet only resulted in a dozen transient events. The 10-meter South Pole Telescope (SPT) has also undertaken a transient survey \citep{Guns2021}, finding 15 transient events associated with 10 objects, eight of which were flaring stars. Additionally, SPT has actively monitored AGN, with variability of one source reported \citep{Hood2023}. Both ACT and SPT are single dish instruments with fairly large ($\sim2\deg$) fields of view; their ability to detect transient events has been limited primarily by their sensitivity. Additionally, both only observe in the millimeter, with no submillimeter bands.

While current generation CMB experiments have had only modest success in cataloging transient events, the work done so far has laid the groundwork for future projects such as AtLAST. The ability of ACT and SPT to detect some millimeter transient events demonstrates the potential for a higher sensitivity, single dish instrument with a large FoV to expand upon the current state of the art. More concretely, these experiments have exposed the need for specialized data products and reduction pipelines to fully explore the transient universe. They have also, however, provided templates for producing just those data products, such as the depth-1 maps used by ACT \citep{Naess2021}. 

In the near future, the 6-meter SO LAT will observe $40\%$ of the sky at three day cadence with several times the depth of current generation CMB experiments, greatly improving on their results. In particular, SO will likely observe a few ($\sim 10$) extragalactic transient events \citep{Eftekhari2022}, up from zero to one such detection by current generation experiments. The 6-meter CCAT has similar expectations \citep{CCAT2023}. SO is being constructed at the same site ACT was located, which is nearby ALMA and both of the sites under consideration in the AtLAST Design Study. CCAT is also nearby, but at a higher altitude allowing for more regular observing at higher frequencies.
Therefore, given the capabilities of these telescopes and the planned survey strategies of SO and CCAT, the observed footprints will overlap completely with the AtLAST accessible sky, enabling excellent synergy between all of them. Transients detected by SO will be reported in near real-time, enabling higher sensitivity follow-up in the submillimeter by AtLAST.

Continuous submillimeter monitoring of nearby star-forming regions has been undertaken for almost a decade by the SCUBA-2 instrument on the 15-meter James Clerk Maxwell Telescope (JCMT). This survey program has resulted in detections of 18 protostars with variable emission \citep{Herzceg+2017, Lee+21, Mairs+2024}. Despite its successes, the JCMT survey is targeted at local star-forming regions, and is therefore generally insensitive to transient events from non-stellar origins. Only a single extragalactic source, an AGN, has thus far been serendipitously detected in these fields \citep{Johnstone+22}. Furthermore, the relatively small field of view size \citep[$\sim8$\,arcmin;][]{Holland2013} makes detection of short-duration bursts unlikely, and to-date only one, a powerful submillimeter flare from a young T\,Tauri star, has been observed \citep{Mairs+19}. CCAT will improve upon the current state of the art, with dedicated monitoring observations of star-forming regions and the Galactic center \citep{CCAT2023}. 

Beyond blind searches for transients and monitoring of a few small fields, a number of current generation millimeter and submillimeter facilities are available to follow-up transients detected at other wavelengths. In particular, follow-up of optically-detected sources and events by the Northern Extended Millimeter Array and ALMA is now routine, with a well established track record of success
\citep[e.g., the follow-up of AT2018cow; see][]{Ho2019, Postigo2018}. There have also been some searches for transient events using ALMA \citep{Mus2022}, although the very limited FoV of ALMA restricts these to only the smallest transient fields (e.g. the Galactic center). 

In addition to interferometric follow-up, the $90$\,GHz MUSTANG-2 receiver on the single-dish 100-meter Green Bank Telescope has demonstrated success in following-up optically detected transients, with \cite{bright2022} highlighting its ``suitability for rapid transient follow-up at high frequencies.'' MUSTANG-2, however, is limited to only operating at $90$\,GHz without polarization sensitivity. Moreover, due to the limitations of the GBT's primary mirror surface accuracy as well as its site ($\approx$840 meters above sea level), no higher frequency instruments can be installed. Finally, the 50-meter Large Millimeter Telescope could potentially be used for transient follow-up in the millimetric bands, particularly with the TolTEC instrument, although to the authors' knowledge LMT has not been so used.
Both the GBT and LMT are also limited in their discovery potential by their few arcminute fields of view (which are factors of hundreds smaller than that of AtLAST).


In addition to transient events, solar system objects such as asteroids, TNOs and comets deserve discussion in this paper. While asteroids and TNOs do not inherently time-evolve, they appear transiently in millimeter and submillimeter surveys, and hence the scanning constraints they place on an instrument and the data products they require are similar to those of true transient events. 

To date the only solar system objects detected by millimeter surveys are asteroids, with both SPT \citep[3 detections;]{Chichura2022} and ACT \citep[$\sim 170$ detections;]{Orlowski-Scherer+23} publishing catalogs of asteroid fluxes. These catalogs are built from stacking of serendipitous observations of asteroids during normal scanning. Future experiments will increase the number of detected asteroids due to the non-linear distribution of asteroid sizes, with SO projected to detect $\gtrsim 1000$ asteroids (Simons Observatory Collaboration, in prep.). 
Moreover, SO may detect a number of TNOs, which to-date have only been detected using very deep targeted observations such as are supplied by ALMA. CCAT will also be primed to commensally detect transient objects, particularly asteroids and trans-Neptunian Objects (TNOs) which are much brighter in the submillimeter than in the millimeter.

In contrast to the above, targeted observations of asteroids and other solar system bodies in the submillimeter is a very well established field, with a long track record of successful observing campaigns \citep{Cordiner2024}. Targeted observations provide key information about asteroid composition which is not accessible in the optical and IR, for example discriminating between metallic and rocky compositions \citep{deKleer+21}. Targeted observations of TNOs have also been undertaken \citep{Lellouch+17}. Still, targeted observation are fundamentally limited in the number of objects that can be observed, which is critical for understanding the formation history of asteroids and TNOs. Full understanding of the taxonomy and composition of asteroids and TNOs will require catalogs of thousands such as only can be compiled by large, sensitive surveys such as those that will be enabled by AtLAST.

We note that AtLAST will not be a dedicated survey facility.  However, it is envisaged that some fraction of its observations will be in survey mode \citep{vanKampen2024,Klaassen2024,Liu2024}, and with a 1-2$^\circ$ FoV, nearly every observation will hold the potential for serendipitous discoveries.

\subsection{State of the Art Deficiencies}
The field of transient millimeter and submillimeter astronomy is still young. As such, there are major deficiencies in the field. Firstly, there are no deep, wide-field surveys. CMB experiments provide the closest approximation, but are limited in depth, possessing sufficient sensitivity to detect only a handful of flaring stars in blind surveys. Moreover, their frequency coverage only extends to the millimeter, although CCAT (up to $850\,\text{GHz}\simeq 0.35$\,mm) will extend this to the submillimeter. SO and CCAT, however, have significantly lower sensitivity than AtLAST. For some science cases this increase in sensitivity can be transformative; the science returns can be much greater than the simple increase in sensitivity would indicate. For example, the asteroid size, and hence flux, distribution steepens significantly below $\sim25$\,km, which corresponds to fluxes accessible by AtLAST but not SO/CCAT (see Section~\ref{sec:ast}).

In the realm of targeted observations, while ALMA and other high sensitivity, high resolution interferometers have provided a wealth of information on transient sources, their semi-automatic triggering means that we are blind to the first few hours of transient evolution. Moreover, simultaneous observations in all spectral bands is currently not enabled with ALMA, restricting science cases which rely on a full spectral understanding of the transient event. For many transient events, such as Fast Blue Optical Transients (FBOTs), understanding the fully spectral time evolution is critical for elucidating the underlying physics, motivating a requirement for an AtLAST receiver capable of simultaneous multi-chroic observations. Moreover, the various observational configurations employed by ALMA mean that their archival data products are very fragmented and inhomogeneous. This makes comparisons between observations difficult, which is of importance for transient and variable science cases. An ideal transient science dataset would be a large legacy data set, with consistent and homogeneous properties over the lifetime of that data set. 

\subsection{The Need for a Larger Aperture, Multi-Beam, Single Dish Submillimeter Facility}
AtLAST's capabilities \citep[see e.g.][]{Klaassen2020, Mroczkowski2024} will enable fundamentally new exploration of transient and variable sources; with its wide FoV and large collecting area $A$, it will be able to quickly survey large regions in search of transients, and can find transients serendipitously in pointed observations and unrelated surveys. This capability is vital both for identifying new transients and for following up transients detected at other wavelengths. AtLAST will also be critical for follow-up of multi-messenger signals that have poor localization, such as black hole mergers and neutrino sources. Additionally, with multiple bands, rapid variability of the SED of sources can be probed without the need for switching of receivers, and calibration can be done much less frequently, meaning that light curves can be produced to study rapid phenomena. Even for state-of-the-art interferometers like ALMA, the calibration overheads (i.e.\ the need to interleave observations of the calibrator and source) can lead to loss of time on source. Finally AtLAST can be used for targeted observations of transients. To do so will require fast response time, high sensitivity, and simultaneous observation in many bands, which can be satisfied by a large FoV, single-dish telescope with an instrument capable of simultaneous, polarized observations in several bands spanning the millimeter and submillimeter ($\sim 90$-850\,GHz).

\textbf{Large Instantaneous Field of View}

AtLAST's wide field of view and high sensitivity will enable it to discover submillimeter transients on a daily basis, both systematically while performing surveys and serendipitously when performing targeted observations. At many other wavelengths, all-sky monitoring has been routine.  In the X-ray, all sky monitors have been nearly continuously in operation since the 1970s, while in the optical, few day cadence of the whole sky to a reasonable depth has been possible for the past decade and a half \cite[e.g.;][]{York2000}. 
In recent years, shallow surveys covering nearly the whole sky have been completed, while moving forward, optical projects like the Argus Array may give deep, high duty cycle coverage in the optical band.  

In the submillimeter, these efforts have been largely absent. With its fast slew speed, wide bandwidth, and large collecting area, AtLAST will be capable of surveying the sky to much fainter limits than even CMB surveys.  For reference, with 50 GHz of bandwidth at 300 GHz, in 2 seconds per position (roughly corresponding to covering the whole visible sky in a day using the bands set forth in \citealt{DiMascolo2024}, see Table~\ref{tab:bands}), AtLAST will reach $\sim 1\,$mJy noise levels, comparable to the full co-added (i.e., several year) SO depth.  If one restricts the daily mapping to a slew covering the inner 2 degrees of the Galactic Plane, then using even 5\% of AtLAST's observing time would reach a noise level of 300 $\mu$Jy every day (see Section~\ref{sec:ded_obs} for a comparison to SCUBA2), with the data collected in a manner that would allow for systematic transient searches and monitoring of longer time-scale variability.

\begin{table}[]
    \centering
    \begin{tabular}{c|c|c}
          ref. freq. & band edges & sensitivity  \\
        GHz & GHz & $\mu \text{Jy beam}^{-1} \text{h}^{-1/2}$ \\
         42. 0 & 30-54 & 6.60\\
         91.5 & 66-117 & 6.46\\
         151.0 & 120-182 & 7.14\\
         217.5 & 183-252 & 9.22\\
         288.5 & 252-325 & 11.91\\
         350.0 & 325-375 & 23.59\\
         403.0 & 384-422 & 39.98 \\
         654.0 & 595-713 & 98.86 \\
         845.5 & 768-905 & 162.51
    \end{tabular}
    \caption{Bands used throught this work. These are based on those used in \cite{DiMascolo2024} and maximize sensitivity given the available atmospheric windows. In general these bands are optimal or close to optimal for transient science, though cases where they differ are noted.}
    \label{tab:bands}
\end{table}

While AtLAST will likely not perform a survey over as large a region as a CMB survey would (i.e. $\gtrsim 40\%$, or several steradians), it will perform regular surveys \citep{vanKampen2024,Klaassen2024,Liu2024} which allow for routine monitoring of known variable sources as well as unbiased blind cataloguing of transient events. Furthermore, deep pointing will allow for serendipitous detection of transients when pursuing other science goals. In both cases the large, fully populated FoV is an essential requirement for any AtLAST design, firstly to survey quickly but secondly to increase the chances of serendipitous transient detections. This capability comes with data processing requirements: maximum cadence maps \citep[such as those used by ACT;][]{Naess2021} need to be made, a real-time alert pipeline needs to be implemented, and known sources such as AGN and dusty variable stars continuously monitored. 

\textbf{Single-Dish, Multi-Chroic}
While the primary science driver for AtLAST for time domain work will come from its wide-field capabilities, for certain science topics, it will exceed ALMA in capabilities for observations involving staring at known point sources.  Specifically, while the ALMA collecting area is about 3$\times$ that of AtLAST, AtLAST can offer multiple frequencies simultaneously and lower duty cycle on calibrators.  It can also offer the ability to make fast measurements which will allow probing stochastic variability, which is challenging for interferometers like ALMA. Additionally, AtLAST will be able to provide accurate localization for bright transients, whether detected in the millimeter band (e.g.\ by CMB surveys) or in other bands or messengers (e.g.\ gamma-ray bursts, or gravitational wave sources), because of its fast slewing, and low overheads relative to interferometers.  

A rapid response to targets of opportunity is critical for understanding fast-evolving transients such as FBOTs, setting another requirement for AtLAST. While millimeter astronomy has traditionally been the province of slowly varying thermal emitters, with greater sensitivity it also becomes an important tool for studying optically thick synchrotron emission. Because opacity for synchrotron tends to reduce as sources evolve and expand, the millimeter band will feature {\it faster} variability than other bands for synchrotron emission, and it becomes crucial to obtain data early and often. As such, obtaining as fast a response time as possible, ideally $\lesssim 1$\,hour, should be a key objective in the development of AtLAST's instrumentation and operating procedures.





\section{Scientific rationale}
There are several Solar System, Galactic, and extragalactic sources that are either expected or are already known to vary at submillimeter wavelengths. However, in general very little is known about the variability of the submillimeter sky. Therefore, the AtLAST transient science case includes both searches for new phenomena and characterization of already identified phenomena. 
As other papers cover the core observational capabilities of AtLAST \citep{Klaassen2020, Mroczkowski2024} or discuss the science cases for AtLAST \citep{Ramasawmy2022, Cordiner2024, DiMascolo2024, Klaassen2024, Lee2024, Liu2024, vanKampen2024, Wedemeyer2024}, this work focuses on the observational constraints imposed by the various 
time-domain science cases. This section will cover the scientific motivation for using AtLAST to study the transient submillimeter sky and some projections for what AtLAST can contribute to this field. After, we consider the specific core capabilities of AtLAST that will enable these studies. The core capabilities we will consider are: (1) surveys of transient fields (e.g., the Galactic Plane and star forming regions), (2) serendipitous transient detection, (3) and pointed observations.  For instance, targetted observations could be tracking stares, dithers, small area Lissajous daisy maps, or those using an instrument-based wobbler 
to probe rapid variability or follow-up transient events from other wavebands.  Solar physics with AtLAST will also be fundamentally variability science, and is covered by \cite{Wedemeyer2024}. Many of the general considerations laid out here are generic to any type of transient and variable source science, but given that such work is only just starting in the millimeter/submillimeter band, we present a broad introduction for a community new to some of these issues.

\subsection{Solar System Science}
Solar System bodies such as asteroids are generally geologically dead, and hence have no intrinsic variability. However, due to their movement across the sky and changes in apparent brightness due to viewing geometry, they appear as variable sources in the context of surveying. 
Moreover, as discussed above, commensal observation, particularly of asteroids, has proven to be a fruitful path for understanding the composition of those bodies \citep{Orlowski-Scherer+23}. Thus, it is critical to understand how AtLAST can contribute to this new and exciting field, and the observational requirements that doing so will place on AtLAST. The planetary science case for AtLAST is discussed further in \cite{Cordiner2024}.

\subsubsection{Asteroids}
\label{sec:ast} As tracers of formation history of the Solar System, understanding the composition of asteroids is of the utmost importance in planetary science. In particular, while the composition of the surfaces of asteroids are relatively well understood from IR measurements \citep[e.g.,][]{Mainzer2011}, the composition of the unconsolidated regolith, or outer layer, is very poorly understood. Millimeter and submillimeter observations of asteroids offer a window into this poorly understood regime, as the regolith is partially transparent at these wavelengths. As such, millimeter and submillimeter observations are sourced up to several centimeters into the regolith. Such observations have generally found a deficit of emission with respect to extrapolations from IR measurements \citep[e.g.,][]{Johnston+82}. In the past, this deficit has been ascribed to a dropping effective emissivity due to scattering within the regolith as compared to the surface \citep{Redman1992}. However, recent observations have suggested that the deficit may instead be due to the temperature in the regolith dropping faster than expected with depth from the surface \citep{Keihm2013}. Systematic observations of asteroids in the millimeter and submillimeter are needed to resolve this issue. Historically, targeted observations of asteroids have been required in the millimeter and submillimeter, which are very costly in terms of telescope usage. Recently, however, CMB survey instruments have reached sufficient map depth to detect asteroids. SPT published the first such asteroid catalog \citep{Chichura2022}, although with only 3 detections they were unable to rule conclusively on the millimeter deficit. ACT followed up SPT with a catalog of $\sim170$ asteroids \citep{Orlowski-Scherer+23}, which did systematically show a deficit of flux as compared to IR extrapolations. Moreover, that deficit had a distinct spectral shape, with a more severe deficit at higher frequencies ($150$ and $220$\,GHz) than at lower ($90$\,GHz). The deficit in the ACT observed asteroids also varied with asteroid class, with S-type asteroids evidencing a stronger deficit than C-types. Both of these observations suggests a physical origin in the regolith composition of the asteroid. 

\begin{figure}[h]
    \centering
    \includegraphics[width=\columnwidth]{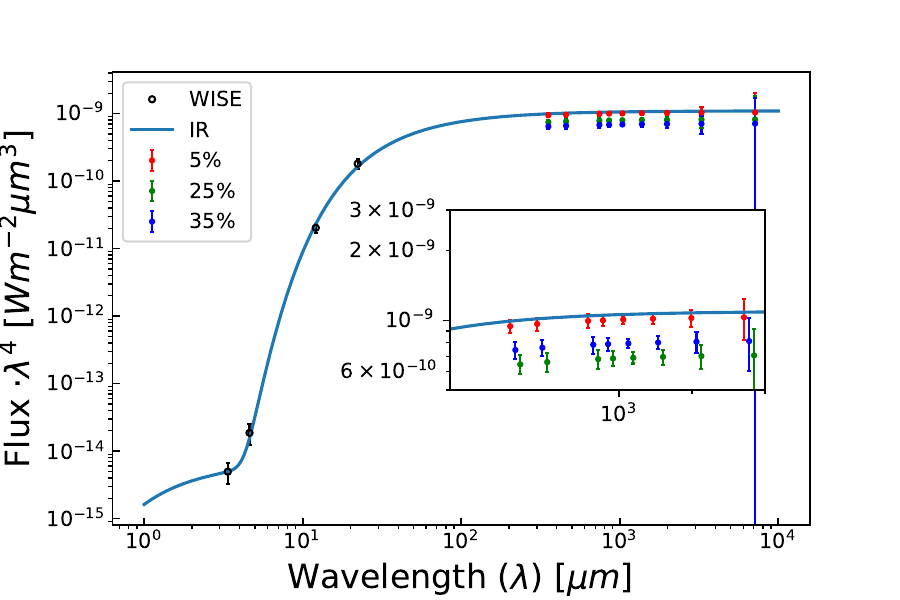}
    \caption{AtLAST projected 1 hour cumulative observations of (4866) Badillo alongside WISE (open circle) data points. The red, green, and blue data points represent projected AtLAST observations reduced by 5, 25, and $35\%$ with respect to best fit IR line (solid blue line), which in turn correspond to the observed deficit at $90$, $150$, and $220$\,GHz by \cite{Orlowski-Scherer+23}, respectively. (4866) Badillo is a 13km asteroid, far smaller than any visible by current generation CMB experiments. AtLAST would enable detection of a submillimeter deficit for even asteroids this small.} 
    \label{fig:atlast-lightcurve}
\end{figure}

The surprising lack of theoretical progress in this field can be ascribed to the lack of systematic surveys of asteroids. 
While the upcoming SO and CCAT experiments will further our understanding in this area, AtLAST is poised to provide higher instantaneous sensitivity and better localization. 
AtLAST would be able to build upon \cite{Orlowski-Scherer+23}, using its broad frequency coverage, fine frequency resolution, and high sensitivity to investigate the origin of the spectral submillimeter flux deficit. In Figure~\ref{fig:atlast-lightcurve} we show a projected observation of (4866) Badillo, a small ($\sim 13$\,km) main-belt asteroid, using AtLAST. In general the observation of asteorids favors broadbands for sensitivity, and so we have chosen frequency bands which match those in \cite{DiMascolo2024} as a nominal case. The projected observation is for one hour, corresponding to $20$ opportunistic observations of $3$ minutes each, which in turn is typical for a year of observations. 
While ALMA can produce similar quality SEDs, it requires targeted observations to do so, which are highly oversubscribed. For AtLAST, on the other hand, results as in Figure~\ref{fig:atlast-lightcurve} would not require targeted observations. Due to AtLAST's large FoV, any observations of the ecliptic can produce opportunistic measurements of asteroid fluxes as they pass through the FoV.

AtLAST will enable submillimeter access to $10$'s of thousands of asteroids as of yet unmeasured in those wavelengths. While the exact distribution of asteroid fluxes in the submillimeter is currently unknown, to first order it will likely follow the cumulative size distribution, which is well described by a power law $\xi(D) \propto D^{-\alpha}$ with $\alpha \simeq -2.5$ \citep{Dohnanyi69}, while the flux falls with the physical radius squared. With one hour observation, AtLAST will reach a sensitivity approximately $100$ times greater than the stacked ACT sensitivity over the same band. Given the likely power law scaling, this corresponds to $10$'s of thousands of asteroids detected. Such a catalog would revolutionize the study of asteroids. 

Of particular note, it is thought that there is a break in the size distribution around $20$\,km \citep{Ishida84,Cellino91}, with the distribution steepening beyond that number. Instruments with sufficient depth to observe asteroids that size and smaller will see significantly more asteroids due to the steepening of the power law, motivating high sensitivity searches in the submillimeter. As shown in Figure~\ref{fig:atlast-lightcurve}, AtLAST will have sufficient sensitivity to probe this population of asteroids; SO and CCAT will not.

Finally, polarization measurements are of interest as they help constrain the metal content of the asteroid's regolith \citep{deKleer+21}. However to date only very sensitivity instruments, particularly ALMA, have been able to detect asteroid polarization. With typical polarization fractions of $\sim1\%$, AtLAST would be able to detect polarization from larger asteroids, provided that the multi-chroic camera was polarization sensitive. 

{\bf Asteroids as submillimeter calibrators:} Asteroids are ideal calibration sources in the millimeter and submillimeter. They are compact bodies with sub-arcsec angular sizes, appearing point-like when observed with submillimeter dishes like AtLAST. Their emission is relatively stable, their trajectories very well known and they are distributed across different local sidereal time ranges. However, only a handful of the brightest ones have been available to the current generation of ground-based observatories like APEX, ASTE, JCMT or SMT, and their light-curves are not well characterized (see \href{https://doi.org/10.26033/b3q7-3b91}{Planetary Data System - Submillimeter lightcurves of asteroids}). There have been efforts to combine space-mission measurements with ground observations to reconstruct the physical and thermal properties of a collection of these asteroids. As a result, submillimeter emission models are available, but their accuracy is still limited as they do not account for daily changes due to rotation or emissivity changes as a function of frequency (see Asteroid-related calibration on the website of the EU-funded Project \href{http://sbnaf.eu/results/bProducts.html}{`Small bodies near and far''}). AtLAST will contribute to the characterization of the light curves of asteroids helping to improve these models, and therefore the suitability of asteroids in general as submillimeter calibrators.

\subsubsection{TNOs}
Additionally, AtLAST would open a new window into the study of other solar system objects in the submillimeter. TNOs are very challenging to observe in the submillimeter, with ALMA and SCUBA providing some number of detections. Due to the very low number of objects that have been observed, the submillimeter flux properties of TNOs are even less well understood \citep{Lellouch+17}. Submillimeter observations of TNOs will be critical for developing a taxonomy for these objects, which is currently very poorly constrained (Bernardinelli, in prep.). Detecting a typical TNO requires submillimeter sensitivity of the order $\sim 100\,\mu$Jy per square arcsecond \citep{Lellouch+17}, which corresponds to roughly 1 hour of observation with AtLAST. As discussed above, this is typical for a year of AtLAST observations, putting the commensal observation of TNOs firmly within reach. 

\subsection{Galactic Science}

\subsubsection{Protostars}\label{section-protostars}

Stars accrete the bulk of their final mass while still embedded in their natal envelope, during the Class~0 and, to some extent, Class~I stage (where the following Class~II and Class~III stages correspond to young stars surrounded by protoplanetary disks and remnant disks and a planetary system, respectively). During the Class~0 phase, the envelope absorbs virtually all light at short wavelengths, i.e., in the optical and near- to mid-infrared regime. Reprocessing of the stellar accretion luminosity, and the surrounding accretion disk emission, to longer wavelengths makes protostars readily observable at far-IR and submillimeter wavelengths \citep{Johnstone+13, Lee+21}.

Several lines of evidence indicate that accretion does not proceed in a smooth, constant fashion, but rather varies episodically, on a range of amplitudes and timescales \citep{Fischer+23}. As one example, large bow-shock systems in protostellar outflows indicate that major accretion events happen every few hundred to few thousand years. In more evolved Class~I and II systems, which can be observed at optical and near-infrared (NIR) wavelengths, variability in accretion tracers is common \citep{Fischer+23}. Luminosity increases by factors of a few up to two orders of magnitude, with durations of weeks to centuries, indicate periods of strongly enhanced accretion. Moving to younger sources, in Spitzer and Herschel multi-epoch infrared imaging of the Orion molecular cloud protostar population, \cite{zakri+2022} identify three outbursts from Class~0 protostars. Based on this sample they suggest that accretion bursts are more frequent and more vigorous the younger the protostar is. They also claim that a significant fraction of mass during the Class~0 phase is accreted during bursts, where the fractions may vary hugely between 3 to 100\% of the final mass. These evolutionary results have been supported by a large study of over one thousand protostars monitored in the mid-IR with NeoWISE \citep{Park+21}.

Nevertheless, only a few of the most deeply embedded and earliest phase Class 0 sources are visible in the near- and mid-IR, demonstrating how incomplete and insecure our knowledge about the fundamental process of star formation still is. To understand how accretion at the earliest phases works, we need a full account of the significance of accretion burst vs.\ steady accretion, i.e., we need to properly quantify the frequency and amplitude of bursts as a function of protostellar mass and evolutionary stage \citep{Fischer+23}.

Wide field multi-epoch submillimeter continuum mapping observations will be the way forward, in particular as there will, for the foreseeable future, not be any facilities capable of doing significantly wide-field far-infrared imaging observations. The submillimeter regime is truly instrumental in characterizing accretion bursts from deeply embedded protostars \citep{Lee+20, Francis+22, Yoon+22}. First, as the majority of a Class~0 protostar's luminosity emerges in the far-infrared to submillimeter regime, protostars are among the brightest sources to be seen at submillimeter wavelengths. Radiative transfer simulations \citep[e.g.][]{MacFarlane+2019a,MacFarlane+2019b} show that from the ground, the shortest submillimeter wavelengths will be best at detecting outbursts. For example, for a given rise in luminosity due to accretion, the corresponding increase in flux will be 3--4 times larger at 350~$\mu$m than at 1.3~mm. Furthermore, multi-wavelength observations will help to break the degeneracy between the amplitude of an outburst and the source geometry \citep{Francis+22}. Both a large telescope and observing at shorter wavelengths, i.e., having finer angular resolution, will also be helpful in detecting outbursts to larger distance: better spatial resolution helps to minimize dilution with the more extended parts of protostellar envelope which are less affected by the central source than by the external ambient radiation field.

Deeply embedded protostellar variability in the submillimeter regime was discovered only fairly recently, and mostly serendipitously,  by \cite{Safron+2015} in the low mass regime and \cite{Hunter+2017} for a high mass protostar. The JCMT Transient Survey \citep{Herzceg+2017,Mairs+2017a,Mairs+2024} constitutes a major step towards a more systematic account of protostellar submillimeter variability. The survey targets, on a monthly cadence, 8 circular fields (diameter 30$^\prime$) in nearby low- to intermediate mass star forming regions (since 2015) and 6 fields at somewhat larger distance on intermediate mass regions. Besides several long-term (months to years) protostellar variables \citep{Lee+20, Lee+21, Yoon+22} the survey has also uncovered the largest stellar flare from a young star, with a variation time-scale of half an hour \citep{Mairs+19}, and more recently detected the first AGN source masquerading as a protostar in a nearby star formation field \citep{Johnstone+22}. The JCMT survey has optimized the calibration process to first reach $\sim2$\% relative uncertainty at 850~$\mu$m \citep{Mairs+2017a}, and more recently achieving $\sim1$\% relative uncertainty \citep{Mairs+2024}, making use of the ubiquitously present emission from extended filamentary structures, cloud cores, and other young stellar objects in the fields. Achieving that level of calibration and maintaining it will be critical for AtLAST to do the best possible science for protostars.

\subsubsection{Stellar flares}
Observations of stellar flares at submillimeter wavelengths are rare 
\citep{MacGregor2018, MacGregor2020}, yet they hold considerable significance for elucidating the underlying physics of this phenomenon, including particle acceleration and plasma heating, as well as for assessing their impact on the habitability of exoplanets \citep[see, e.g.,][and references therein]{Lammer2003,Segura2010,Konings2022,Bellotti2023}. 
The submillimeter flux density of detected stellar flares tends to exceed that of the quiescent flux by factors ranging from 10 to 1000. Assuming a typical chromospheric temperature during quiescence of approximately $10^4$\,K (at 100\,GHz) as observed in cool stars \citep{Mohan2022}, the flux density during a flare can span from approximately 0.5 to 50\,mJy for a star situated at a distance of 10\,pc.

Apart from main sequence stars, 
pre-main-sequence (PMS) stars, which rotate fast and have strong (kGauss) magnetic fields, are known to be very magnetically active. The extent to which this activity manifests at submillimeter wavelengths and how it relates to corresponding activity at centimeter radio wavelengths and in the X-ray regime is poorly explored. The few millimeter flaring objects detected to date either result from serendipitous discoveries \citep[e.g.,][]{Bower+2003,Bennettlovell+2024} or from limited monitoring campaigns such as the JCMT transient survey (see Sect.~\ref{section-protostars}) or the ALMA monitoring study of part of the Orion Nebula Cluster of \cite{Vargasgonzalez+2023}. Detected flare amplitudes range from mJy to few hundred mJy (at typical distances of a few hundred pc), while their timescales range from seconds to days \citep[e.g.,][]{Salter+2008, Vargasgonzalez+2023}. Besides scaled-up solar-type flaring, interaction of the magnetospheres in binaries is thought to contribute to PMS star flaring activity \citep{Salter+2008, Massi+2008, Mairs+19}. It is not at all clear whether in very young objects, which are surrounded by massive disks, interactions between the (proto)stellar and disk magnetospheres could provide an additional means to produce strong flares \citep[e.g.,][]{Takasao+2019}, in particular as the very youngest sources are also obscured at X-ray wavelengths, blocking an otherwise well established channel to detect magnetic flaring events.

It is essential to underscore that the flux density exhibits variations over timescales ranging from a few seconds to minutes. Observing the temporal evolution of a flare, which is imperative for discerning the underlying physical mechanisms, necessitates datasets with  correspondingly high temporal cadence. Such datasets are typically not obtained through serendipitous detection or as part of a survey but demand dedicated sit-and-stare observations of the same region.

AtLAST would facilitate such sit-and-stare observations for time-resolved light curves for stellar flares. Leveraging the high sensitivity of AtLAST, it should be feasible to detect stellar flares in cool main sequence stars at distances of up to approximately $\sim$40\,pc with integration times on the order of a few seconds to a minute, thereby resolving the relevant timescales. Moreover, the anticipated distinctive ability to conduct simultaneous observations at multiple frequencies, a feature typically absent in other submillimeter instruments and surveys, especially when coupled with polarization capabilities, would empower AtLAST to employ critical diagnostics such as the modulation of the millimeter spectral index, serving as an indicator of stellar atmospheric activity \citep{Mohan2022}, and providing stringent constraints on the emission mechanisms and estimations of magnetic field strengths in flaring active regions.

The full potential of flare observations at millimeter and submillimeter wavelengths is realized through coordinated, strictly co-simultaneous observations with complementary telescopes spanning as broad a wavelength range across the electromagnetic spectrum as feasible. Once the Square Kilometer Array (\citealp[SKA;][]{Dewdney2009}) starts operations in the next decade in the radio band, it will facilitate coronal tomographic observation by virtue of its sensitive wideband radio imaging capability. AtLAST will be able to complement the observations with chromospheric tomography using multi-waveband millimeter monitoring, thus substantially enhancing the diagnostic potential of such combined data sets. 

Such coordinated endeavors, increasingly prevalent in multi-messenger astronomy and customary for solar observations, necessitate PI-led operations and coordination between observatories starting with policy development and proceeding through the proposal acceptance stages. These policies will be developed as part of the science operations planning. The paucity of such coordinated stellar observing campaigns so far constitutes a primary obstacle hindering the confident detection of stellar coronal mass ejections (CMEs) -- a pivotal constituent of exo-space weather. The efficacy of such campaigns would be significantly augmented by AtLAST's expansive FoV and high angular resolution, enabling the simultaneous observation of multiple stellar targets while separating the flare source(s) from other elements such as nearby active companion stars or disks, thus rendering AtLAST well-suited for time-resolved coordinated observations of stellar flares.
Comparing such observed stellar flares to their solar analogues, which could also be observed with AtLAST 
\citep[see][for the solar science cases]{Wedemeyer2024}, would enhance the interpretation and understanding of the physical mechanisms driving this phenomenon with essential implications of solar/stellar activity and the impact on exoplanet habitability. In addition, the observations of stars described here will simultaneously provide us with observations of their circumstellar disks, helping us to build up a complete picture of these exoplanetary systems \citep[for more details on the disk observations see][]{Klaassen2024}.

\subsubsection{Red Novae}
Outbursts of classical novae that are powered by thermonuclear runaway explosion on the surface of a white dwarf, also occasionally produce submillimeter transients owing to enhanced free-free emission that extends out to radio wavelengths \citep[for a review, see][]{Seaquist_2008}. Bright events display millimeter-wave fluxes at a level of a few mJy \citep[cf.][]{KamiATel}.  AtLAST should be able to serendipitously catch such hot outbursts at a rate of a few events per year. These observations would be useful in reconstructing the full spectral energy distributions (SED) of individual classical novae, which, in turn, should shed more light on the ejecta evolution in these evolved binaries.

Red novae are a group of Galactic and extragalactic transients powered by stellar collisions that take place in binaries with a noncompact component. Because their energy comes predominantly from accretion and not from nuclear fusion like in classical novae \citep[but see][]{Tylenda2024}, red novae cool down to low temperatures ($\sim$2000 K) after the eruption. Consequently, the material dispersed during the collision quickly recombines, forms molecules, and effectively condenses into dust \citep{Kami2018}. Depending on the progenitor, the type of dust produced is carbonaceous or (more commonly) silicate, but red-nova remnants have also displayed remarkably strong signatures of alumina dust \citep{Banerjee2015}. The formation of dust is observed in real time \citep[e.g.][]{Wisniewski, Nicholls}, similar to what has been observed in dust-forming classical novae. Since a merger product is expected to have strong magnetic fields \citep{Schneider}, the submillimeter emission of dust can be strongly polarized. AtLAST surveys have a good chance of catching those dusty transients, and dedicated AtLAST polarimetry may provide terms for a verification of their enhanced magnetic fields.

Red novae are a manifestation of collisions in binary systems at different evolutionary stages, including protostars, main-sequence stars, subgiants, and red giants. It is expected that mergers in the more evolved systems, whose extended stars have less gravitationally bounded envelopes (e.g. red giant branch, RGB, or asymptotic giant branch, AGB, stars) produce especially copious amounts of dust \citep{TylendaBLG, MacLeod} that can give strong signatures in millimeter and submillimeter continuum. Known examples are red novae CK Vul and BLG-360, whose remnants are observed at $\sim$10--100 mJy. The dust mass in these remnants is estimated to be on the order of 0.1\,M$_{\odot}$ \citep[][Steinmetz in prep.]{Kami2021} and both most likely had RGB progenitors. Since compact binaries are likely to interact after leaving the main sequence (Case A), these are the most likely systems to be observed as red novae and would be primary detection targets for AtLAST.

The current rate of red novae that are observable in the visual and NIR is estimated at 2--3 bright events per decade per galaxy \citep{Kochanek, Howitt}, but no event of this type has been identified in the Milky Way since 2008. There are theoretical and observational hints that some red novae and similar objects become so highly embedded in dust that they appear as bright transients mainly at infrared wavelengths \citep[for example, SPIRITS of][]{Jencson}. Since little has been done in monitoring the sky in the infrared and at longer wavelengths, we are likely missing a substantial population of red novae, especially the outbursts from the more extended giant progenitors. An AtLAST survey extending to millimeter wavelengths can serendipitously catch some of these Galactic transients during and just after the outburst. Predicting these events has proven to be notoriously difficult, \citep{Pietrukowicz}. The duration of the outburst can vary from months to years, so a survey cadence of a few weeks would suffice to not miss any bright millimeter-wave event. Even if they turn out to be rare, detections of red nova outbursts at millimeter and submillimeter wavelengths would be important for revealing the most dust-obscured stellar collisions. This will substantially inform stellar population synthesis and deepen our understanding of binary evolution, especially when common-envelope evolution is concerned.   

Admittedly, it will be hard at first to distinguish at millimeter wavelengths alone a red nova from a short-lived accretion burst in a protostar, as luminosities, dust temperatures, and time scales are comparable in both types of events (cf. Sect.~\ref{section-protostars}). However, obtaining complementary follow-up observations at shorter wavelengths can often uniquely settle the identification. 

\subsubsection{X-ray binaries (XRBs)} 

Multi-wavelength studies of XRBs have shown that a significant fraction of the dissipated accretion power may be released in the form of relativistic jets. Relativistic jets are common to accreting compact objects on all mass scales, from neutron stars or stellar-mass black holes in XRBs to supermassive black holes in AGN. The variability timescales in XRBs, from milliseconds to days and months, offer a distinct advantage for jet studies, allowing us to witness jets being launched and quenched in real time. This has led to an association of the presence of jets with a radiatively inefficient accretion state, also characterised by the Comptonised emission observed in hard X-rays and therefore called ``hard-state'' \citep[see e.g.;][]{Tananbaum1972, Fender2014}. However, what triggers the quenching and re-ignition of jets remains unknown.

In the past decade quasi-simultaneous observations from radio to X-rays with a cadence of a few days during XRB accretion outbursts have revealed significant changes in the jets that add to the known changes in the accretion flow and occasionally detected relativistic ejections \citep{Fender2004, Migliari2006}, also observed at millimeter wavelengths \citep{Tetarenko2017}. Specifically, the jet spectral break from optically thick to thin emission, that lies at optical-infrared wavelengths during hard-states, peaks at submillimeter wavelengths during accretion state transitions to further move to longer wavelengths before jet emission is quenched \citep{Russell2014, DiazTrigo2018}. Conversely, the break moves back to submillimeter wavelengths during jet re-ignition \citep{Corbel2013}. Unfortunately, these studies are limited to a handful and suffer from sparse sampling especially at submillimeter wavelengths. AtLAST will be capable of providing multi-band observations with daily cadence for a few days around the state transitions to combine with multi-wavelength coverage from radio to X-rays and precisely determine the evolution of the jet broadband spectra, thus enabling systematic studies of transient and persistent neutron stars and black holes and shedding further light in the jet launching and quenching mechanisms.

An extreme example of the type of phenomenology that requires observations to be as continuous as possible is understanding the response of relativistic jets in neutron star X-ray binaries to thermonuclear bursts on the neutron star surfaces, which can be used both to map out the speeds of jets launched near neutron stars, and to make a fundamental understanding of the manner in which jets are launched by accretion disks by studying the process under circumstances of extreme variability. In \cite{Russell2024}, it was found that jets brighten in the radio band in response to bursts, and that the response is faster and more intense at higher radio frequency.  With AtLAST, such work could be done significantly better than in the radio band, with the submillimeter bursts standing out more against the stochastic jet variability, and with the time lags between bands easier to detect.  On the other hand, since the bursts can be expected to be only tens of seconds long in the submillimeter band, interruptions for phase calibration could cause bursts to be missed entirely, leading to requirements of much longer exposure times.

In addition, for the brightest sources, fast-timing studies in the millimeter band alone could become possible. Fast-timing multi-wavelength studies of black hole outbursts have revealed that emission is highly correlated between different bands, measuring time-lags ranging from hundreds of milliseconds between the X-ray/optical bands to minutes between the radio/submillimeter band \citep{Tetarenko2021}. Modelling of this variability yielded precise measurements of the speed and collimation angle of the black hole XRB MAXI J1820+070 and highlighted the “need to simultaneously sample more closely spaced frequency bands in between the radio and millimeter regimes (30–100 GHz)” \citep{Tetarenko2021}. To contribute to this field would require simultaneous multi-band observations spanning at least one order of magnitude in frequency and with a sensitivity of $\sim 100\,\mu$Jy (30 sigma) in less than 10 seconds between 35 and 100 GHz, thus sufficient to explore the variability timescales of a few seconds expected in the millimeter regime in the Fourier space \citep{Maccarone2019}. In particular, two bands are necessary to probe jets speeds, while the ideal camera would have at least three bands, enabling constraints on the jet acceleration. This therefore sets a requirement for AtLAST to have at least one receiver capable of multi-chroic observations in at least 3 bands spanning $30$-$850$\,GHz; for any reasonable bandwidths set forth, the sensitivity requirement will be achievable. 

\begin{figure}[!h]
    \centering
    \includegraphics[width=\columnwidth]{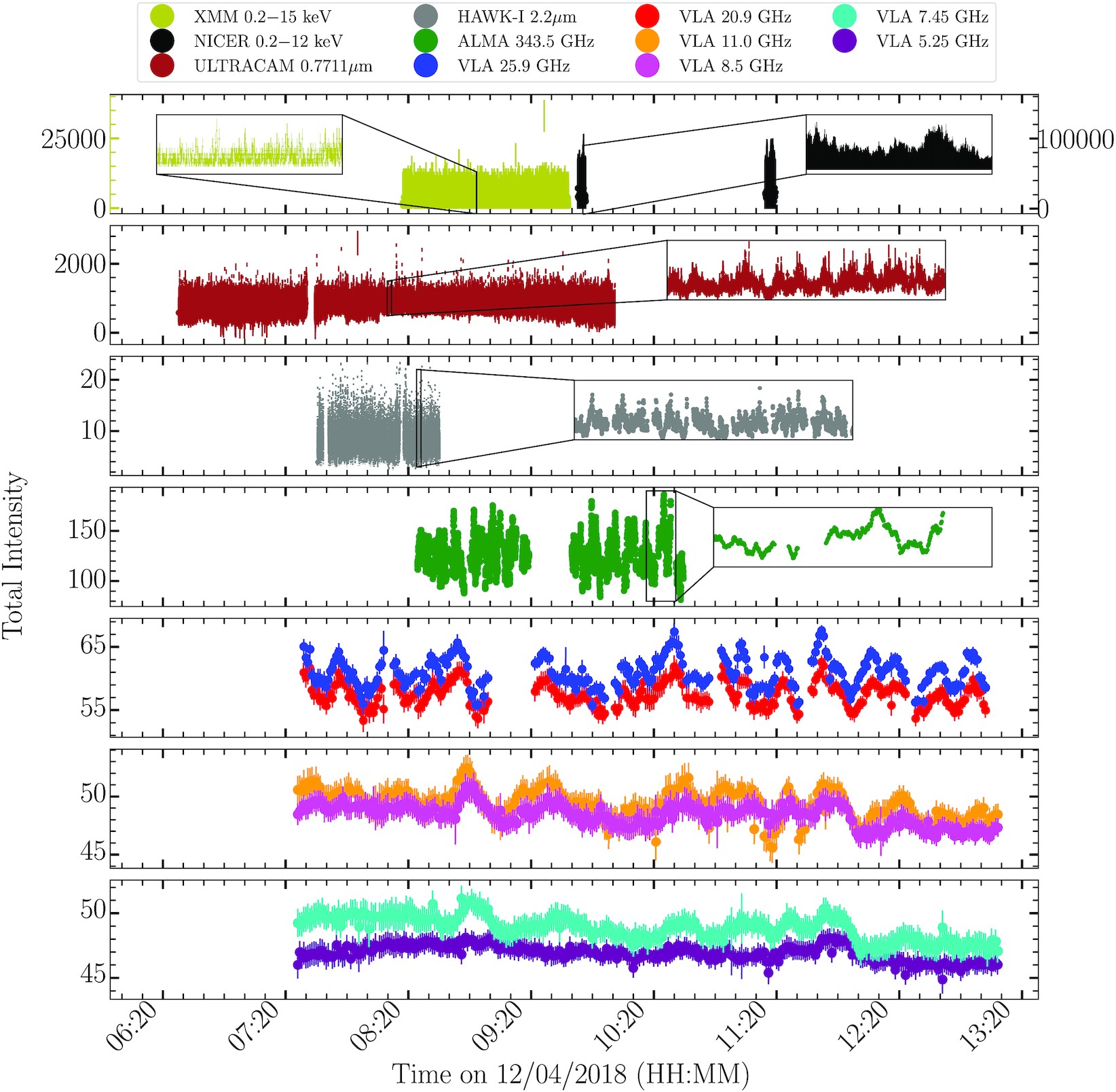}
    \caption{A multiwavelength campaign on the black hole X-ray binary MAXI J1820+070 in outburst, from \citealt{Tetarenko2021}.  The emission from infrared through radio is thought to be synchrotron from a relativistic jet.  The increased smoothness of the light curves as a function of increasing wavelength results from the light travel times across the jet, while the lags of longer wavelengths behind shorter wavelengths can be used to map jet speeds along the jet.  AtLAST's capability of making high sensitivity, multi-chroic measurements allow mapping out the jets' speeds and structures.  } 
    \label{fig:xray-binary}
\end{figure}

\subsection{Extragalactic Science}

\subsubsection{Gamma-Ray Bursts (GRBs)}
Jets are ubiquitous in astrophysical sources, from young stellar objects to binaries in our Milky Way to blazars and tidal disruption events \citep[e.g.][]{Peer+2014}. Yet, a conclusive picture of the jet launching mechanism and the jet structure is still missing, with different models giving rise to different degrees of magnetization, outflow geometries and emission profiles. Since scaling relations have been found between jets from various source classes \citep{HeinzMerloni2013}, the leading acceleration and radiative mechanisms inside jets are believed to be the same. Given the restricted
number of parameters accessible for each source , the large unconstrained parameter space prevents rapid theoretical progress. Although not spatially resolved, GRBs offer the crucial advantage of rapid temporal evolution providing a promising route to understand jet physics.

The dissipative (shock) fireball model, the basic scenario for the emission process of GRBs \citep{MeszarosRees1993,MeszarosRees1997}, assumes a very large energy deposition inside a very small volume, constrained by causality and the observed variability timescales of GRBs to about $<$100 km. This leads to an optically thick, highly super-Eddington $\gamma$e$^{\pm}$ fireball which 
converts most of its radiation energy into kinetic energy, i.e., bulk motion of a relativistically expanding blast wave (Lorentz factors $\Gamma \sim$ 10$^{2-3}$). When the blast wave interacts with the circum-burst medium, an external shock is formed, the macroscopic properties of which are well understood. Under the implicit assumptions that the electrons are ``Fermi'' accelerated at the relativistic shocks, and that they have a power-law distribution with an index $p$ upon acceleration, their dynamics can be expressed by 4 parameters: (1) the total internal energy in the shocked region as released in the explosion, (2) the density $n$ and  radial profile of the surrounding medium, (3) the fraction of shock energy that goes into accelerating electrons $\epsilon_{e}$, and (4) ratio of the magnetic field energy density to the total thermal energy,  $\epsilon_B$. 
According to standard synchrotron theory for a continuous injection of electrons, which is the case for ongoing plowing of the forward-shock into the ISM, 
the arising synchrotron spectrum is described by a four-segment power law, separated by the cooling frequency $\nu_c$, the typical synchrotron frequency $\nu_m$, corresponding to the lower energy limit of the accelerated electrons,
and the self-absorption frequency, $\nu_a$
\citep{Meszaros+1998,Sari+1998,GranotSari2002}. 

Measuring the characteristic breaks in the SED, the spectral indices in between, and their temporal evolution allows to uniquely determine the jet and fireball parameters, such as the jet inclination and opening angle, the ambient medium's density structure, and most notably the energy of the explosion and the energy partition between electrons and magnetic field ($\epsilon_e$/$\epsilon_B$).
An interesting open question is that of a correlation (predicted from Weibel shock theory \citep{Medvedev2006}) or anti-correlation of  $\epsilon_{\rm e}$  and $ \epsilon_{\rm B}$ (as expected from radiative efficiency arguments), while observations also hint at a changing $\epsilon_{\rm B}$ throughout a GRB afterglow \citep{Filgas+2011}.

\begin{figure*}[t]
    \centering
    \includegraphics[width=\textwidth]{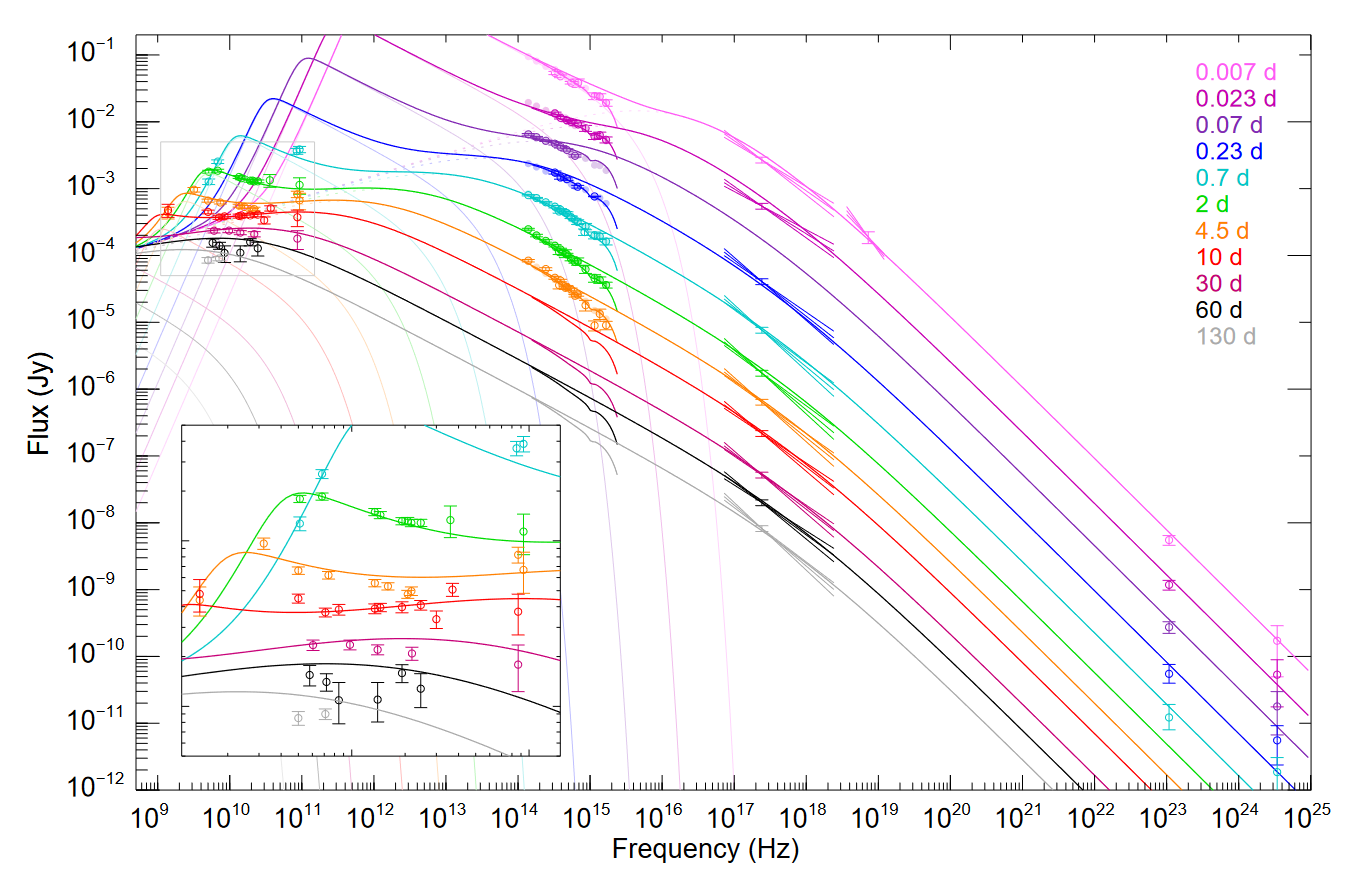}
    \caption{Afterglow of GRB 130427A showing data (open points with errorbars) as well as the best fit model (lines). Colors correspond to time after the GRB emission. Note the very rapid evolution of the peak associated with $\nu_m$ from $10^{12}$ to $10^{10}$\,Hz in 2 days, with much of the evolution occuring in the first few hours. AtLAST would exactly cover the frequencies of interest. While the AtLAST points are not plotted, the typical uncertainty would be $\lesssim 10^{-4}$\,Jy. Figure reproduced from \cite{Perley2014}. } 
    \label{fig:GRB}
\end{figure*}

To achieve these goals, it is crucial to sample the {\em full SED} --- and its temporal evolution --- from the radio to the X-ray regime as completely as possible, starting as early as possible. 
In particular the synchrotron frequency $\nu_m$ falls in (or better, sweeps through) the submillimeter to radio regime. In Figure~\ref{fig:GRB} we have reproduced Figure~10 from \cite{Perley2014}, which shows the rapid spectral evolution of peak associated with $\nu_m$, which falls from $\sim10^{12}$ to $10^{10}$\,Hz in 2 days, with particularly rapid evolution over the first few hours. 
As $\nu_m$ marks the peak of the SED, measuring the GRB brightness in this regime is also important to constrain its overall luminosity. It should be noted that submillimeter observations are preferred over radio observations since the latter may suffer from strong scintillation effects during the first days \citep[see][for an extreme case]{Greiner+2018}. Moreover, the submillimeter regime is also particularly useful to test for deviations from the standard forward-shock fireball model, requiring, e.g., an additional reverse-shock (marking the region where the jet is decelerated to the velocity of the forward-shock), or even more complicated configurations, as is argued (e.g.) to be necessary to explain the submillimeter observations of the GRB 221009A (the "Brightest Of All Times - BOAT") afterglow by \cite{Laskar+2023}.

To date, however, the millimeter to submillimeter regime has been covered only relatively sparsely due to the lack of suitable facilities, which are able to provide sensitivity, broad wavelength coverage, and rapid response.
Observations at X-rays are typically done within minutes, so far with Swift, and in the future with SVOM, the Space-based multi-band astronomical Variable Objects Monitor \citep{SVOM2019}. Robotic optical/NIR telescopes have similar reaction times, and Target of Opportunity (ToO) observations at 8\,m class telescopes are typically done within an hour after a GRB. Submillimeter observations starting at this timescale (1 hour or earlier) after a GRB are deemed sufficient for the above described observational tests of the standard forward-shock model. Even shorter reaction times will be required to probe 
the reverse shock, seen
as an optical flash 50~s after the gamma-ray detection of GRB 990123, followed by a 8.4~GHz radio detection 19~h later \cite{SariPiran1999}; with $\nu_m \propto T^{-73/48}$, $\nu_m$ would reach the submillimeter regime just one hour after the GRB. 
Thus, it is obviously highly desirable to provide reaction times substantially lower than one hour in order to catch the rise and fall of corresponding flashes in the submillimeter regime.

AtLAST will be uniquely positioned to probe the emission of GRBs in the submillimeter to radio regime on sub-hour timescales. A rapid response protocol must be developed for automatic triggering of ToO observations (see Section~\ref{sec:ded_obs}) which will be perfectly suited to following-up GRBs. Moreover, multi-epoch observations over the following few days to weeks will be important to follow the temporal evolution of the GRB afterglow. In addition to requiring a response time shorter than one hour, the GRB science case requires continuum instruments providing superb instantaneous sensitivity (i.e., wide bandwidth, limited mostly by the atmospheric transmission windows) and ideally being able to observe simultaneously cover multiple frequency bands (e.g. with a multi-chroic camera). This will provide simultaneous photometry over the full millimeter to submillimeter regime and short-cadence multi-band lightcurves particularly during the early phase of the GRB follow-up. 

\subsubsection{Fast Blue Optical Transients (FBOTs)}
\label{sec:fbots}
The advent of all-sky and nightly observing cadence optical transient surveys, such as the Zwicky Transient Facility (\citealt{Mairs+19}) and the Asteroid Terrestrial-impact Last Alert System (\citealt{tonry2018}) has driven the discovery of new classes of transient events. Among these discoveries are optical transients that evolve quickly (days timescales), are spectrally blue, and cannot be powered by the decay of $^{56}\rm{Ni}$; the FBOTs. Studies of the multiwavelength SEDs of FBOTs indicate that a newly formed compact object likely powers the synchrotron afterglow, and radio/submillimeter observations reveal that the outflow from the explosion is semi-relativistic and faster than in traditional supernovae \citep[e.g.][]{bright2022,ho2022,coppejans2020,margutti2019}. Due to their intrinsic luminosity FBOTs can be observed out to significant redshifts (e.g. $z=0.243$ for AT2020xnd \citep{bright2022,ho2022}. Submillimeter observations of FBOTs are currently sparse, with only a handful of sources detected \citep[e.g.][]{margutti2019,Ho2019,bright2022}, however such observations are essential in constraining the early-time behaviour of the outflow as it interacts with stellar material lost in the final year of the progenitor star's life. As the outflow expands, the peak of the radio/submillimeter SED moves to lower frequencies and the peak flux density drops, which indicates that rapid follow-up (minutes to hours to days post discovery) is a particularly desirable capability. With high sensitivity (capable of pushing well below $1\,\rm{mJy}$ in an hour) and simultaneous observing frequencies, AtLAST has the potential to revolutionise the follow-up of FBOTs through monitoring the evolving submillimeter SED during the early phases of these cosmic explosions, and consequently allow us to better understand the properties of FBOT outflows. Additionally, it is possible that a small number of FBOTs could be discovered serendipitously as part of regular survey observing (e.g. \citealt{Eftekhari2022}), with the exact number uncertain due to the sparsity of submillimeter observations at appropriately early times and depending on the survey strategies applied on AtLAST.

\subsubsection{Active Galactic Nuclei (AGN)} 
\begin{figure*}[t]
    \centering
    \includegraphics[width=0.48\textwidth]{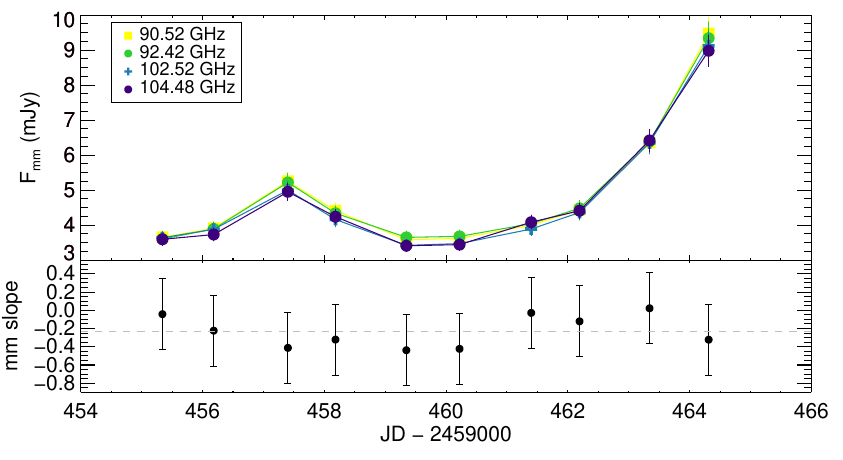}
    \includegraphics[width=0.42\textwidth]{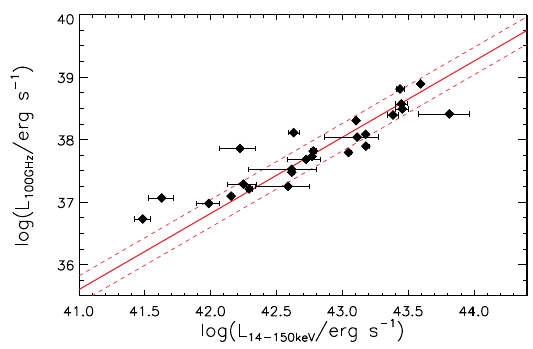}
    \caption{{\it Left panel:} Millimeter light curve of IC\,4329A during a 10-days dedicated observational campaign in four different frequencies \citep{Shablovinskaya2024}. The bottom panel shows the millimeter spectral slope, with the grey dashed line illustrating median value ($-0.23$). {\it Right panel}:  100\,GHz continuum luminosity versus the intrinsic 14--150\,keV luminosity for the sources from the volume-limited hard X-ray selected sample of \citet{Ricci2023}. The solid and dashed red lines represent the best fit to the data and the 1$\sigma$ scatter (0.22~dex), respectively. } 
    \label{fig:AGNvariabilityFig}
\end{figure*}

AGN show flux and spectral variability across the whole electromagnetic spectrum. Such variability can be a powerful probe into the complex environments surrounding accreting supermassive black holes at the centers of galaxies. Submillimeter and millimeter-wave emission from AGN is often neglected because of the assumption that this band is dominated by non-variable large-scale star formation. However, theoretical studies have argued that the $\sim 100-200$\,GHz continuum emission observed in AGN could be associated with self-absorbed synchrotron emission produced by electrons located in the X-ray corona at a few gravitational radii from the supermassive black hole \citep{Laor2008,Inoue2014}. 
Observational studies carried out over the past several years have found results in agreement with this idea. In particular: (1) rapid variability in the millimeter band (a few days, \citealt{Baldi+15} at 95~GHz), which suggests that the scales are very small ($<0.01$~pc, \citealt{Behar+20}). A recent 10-day monitoring of the radio-quiet AGN IC\,4329A carried out by ALMA has shown very rapid and strong variability at 100\,GHz, up to a factor $\sim 3$ \citep[left panel of Fig.\,\ref{fig:AGNvariabilityFig};][]{Shablovinskaya2024}; 
(2) an excess in the SED at 100-300 GHz, exceeding the extrapolation from lower frequencies ($<10$~GHz, \citealt{Inoue+Doi18}); 
(3) the almost ubiquitous presence of an unresolved core at millimeter wavelengths in AGN from high resolution ($<100$\,mas) ALMA observations of a volume-limited sample \citep{Ricci2023};
(4) a very tight correlation between the hard X-ray emission and the nuclear millimeter (100-230 GHz) emission \citep{Kawamuro+22,Kawamuro2023,Ricci2023}. Indeed, a recent study at 100 GHz found a scatter of only 0.2~dex between hard X-ray and millimetre emission at 100 GHz (right panel of Fig.\,\ref{fig:AGNvariabilityFig}, \citealp{Ricci2023}), which suggests a potential new way to detect highly obscured AGN activity. Radiation at 100-200\,GHz can in fact reach column densities $\sim 3$ orders of magnitudes higher than the current limit of X-ray observations ($N_{\rm H}\sim10^{24}$~cm$^{-2}$), allowing detection of emission from AGN obscured by extreme column densities, up to $N_{\rm H}\sim10^{27}$~cm$^{-2}$. 

\begin{figure*}[h]
    \centering
\includegraphics[width=\textwidth]{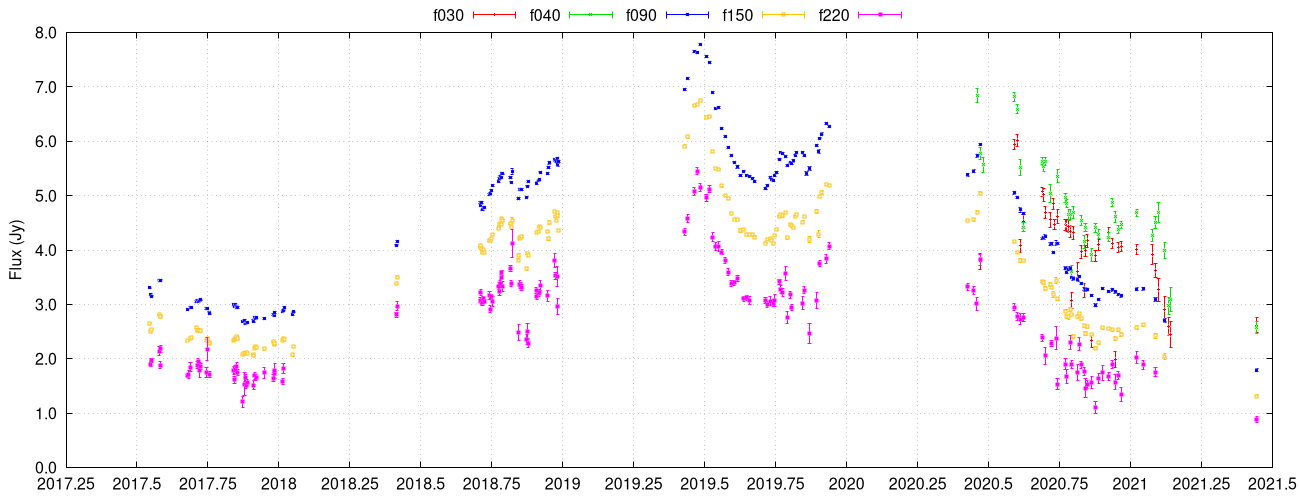}
    \caption{Light curve of QSO~B0208-512 as observed by the Atacama Cosmology Telescope (ACT), covering 30-230 GHz over four years. The source is a BL Lac at $z\sim1$ and shows variations on short timescales (days), due to its relativistic jet pointing towards us. ACT measured thousands of such light curves, though most with lower signal-to-noise ratio than this one. AtLAST's 69.4 times as large collecting area results in 69.4 times higher integration speed, meaning that AtLAST could reach the same sensitivity in 1.4\% of the time.}
    \label{fig:act-lightcurve}
\end{figure*}

Submillimeter variability can be a great additional probe of AGN activity, allowing identification of AGN that would be otherwise missed by X-rays or optical surveys because they are heavily obscured. The study of AGN variability in the millimeter and submillimeter has been limited, with some results from CMB experiments (see Figure~\ref{fig:act-lightcurve}), a few ALMA calibrators \citep{Bonato2018}, and one blazar found as an interloper within a nearby star-forming region \citep{Johnstone+22}. A wide-field telescope capable of daily cadence observations, such as AtLAST has the potential to survey hundreds of AGN and identify highly obscured AGN, potentially opening a new window to study the most obscured accretion events in the Universe.

\subsection{Coalescing Stellar Black Hole/Neutron star Binaries}

Among the most exciting astrophysical moments of the last decades was the detection of gravitational waves (GW) from (stellar mass) compact object mergers \citep{Abbott+2016} and their connection to corresponding phenomena observed by electromagnetic waves \citep[e.g.,][]{Abbott+2017}, constituting the beginning of the era of multi-messenger astronomy. While in particular the electromagnetic observations of GW170817 were mostly triggered by the GW event (with the exception of the detection of the corresponding GRB), ongoing improvements in sensitivity and data processing of the GW detector facilities promise the delivery of pre-merger triggers, as it will become possible to detect and announce the GW signal already during the inspiraling phase of the merging bodies \citep[e.g.,][]{Magee+2021,Yu+2021,ChatterjeeWen2023}. Presently pre-warnings are possible only a few to few ten seconds before the merger, with poor localization (thousands to hundreds of square degrees). While further improvements arguably will be possible with currently operating facilities, major steps forward can be expected with the advent of the next generation of GW facilities (e.g., the Einstein Telescope): pre-warning times of $\sim$10~min to few hours with a localization on the order of 100~square degrees to few ten square degrees should be possible, where the localization would get more precise closer to the actual merger \citep[e.g.,][]{Miller+2024, HuVeitch2023}. AtLAST will require a large FoV to explore the relatively large area of localization in a timely manner.

In terms of physics, merging compact objects are amongst the most extreme laboratories accessible to observations \citep[e.g.][]{Baiotti2017, Corsi+2024}. To fully understand the processes happening during and after the merger, it is crucial to understand the properties of the participating objects before the merger. If at least one of the compact objects is magnetized, a plethora of phenomena can be expected to be observable before the merger due to the interaction of the magnetosphere with the other object or the other objects magnetosphere \citep[e.g.,][]{Metzger2016, Most2023}. Reconnection of antiparallel fields in the space between two neutron stars with antiparallel fields is inevitable as the two objects approach each other; after reconnection, the magnetic poles of the objects are connected, further inspiral will wind these fields up, giving rise to further reconnection, etc.\ \citep[e.g.,][]{Palenzuela+2013}. (Gyro)synchrotron emission from these reconnection events (or from any charged particles moving in these extreme magnetospheres) may provide a unique opportunity to constrain magnetic fields in these objects pre-merger, and provide critical input in assessing the role of magnetic fields during  and after the merger itself. To achieve this an instrument with a wide field-of-view, high sensitivity, polarized detectors, and the capability to respond rapidly to triggers will be the key.

\subsection{Massive Black Hole Binaries (MBHBs)}

Massive black hole (MBH, $M_{\rm BH}>10^{5} M_\odot$) binaries are greatly important for understanding hierarchical structure formation theories, the growth and demographics of supermassive black holes (SMBHs), and the physics of accretion and feedback. Even more compelling,  in the final stages of their journey towards coalescence, these systems are predicted to be amongst the loudest emitters of gravitational waves in the low-frequency ranges, close  to the frequencies where some of the current and incoming gravitational wave experiments are most sensitive (e.g., PTA - \citealt{Hobbs+10}; LISA - \citealt{Amaro=Seoane+23},  LGWA - \citealt{Harms+21}). Following the $\Lambda$CDM cosmological predictions, galaxies grow hierarchically through mergers, and the MBHs hosted in their nuclei sink to the center of the merger remnant because of dynamical friction. Thus, MBHBs, stalling for a significant time ($\sim$1 Gyr) at sub-pc separation, should be common systems in the Universe. In spite of their expected large number, they are observationally rare systems. 

The  surrounding  gas  can  accrete onto each of the MBHs \citep{Dotti+07}; in this case, an active MBH (AGN) binary  system should be present. Angular resolution required to directly detect binary AGN is well beyond the capability of most current facilities, except for the local Universe by using very high-resolution interferometric observations \citep[VLBI;][]{EHTI,EHTII}. Therefore only indirect methods are feasible to unveil large samples of possible candidates. Due to the orbital motion of the two BHs, periodic modulations are expected, with periods comparable to the binary period. Depending on the individual masses, mass ratio, and relative separations, periods could vary from years (when the BHs become bound to each other) to minutes (final in-spiral phase). Due to the decreasing residence time with decreasing separations \citep{Gualandris+22}, MBHBs could be rare sources from the observational point of view. 
Additionally, MBHs binaries are expected to be embedded in a large amount of dust and gas, thus they would appear strongly obscured and elusive both in the UV and optical bands. To date, although hundreds of candidates have been found, all with masses higher than 10$^7$~M$_{\odot}$, \citep[see e.g.,][and references therein]{Valtonen08, Valtonen+Pihajoki13, Shen+13, Runnoe+17, Severgnini+18, Li+19, Liu+19, Serafinelli+20, Chen+20, O'Neill+20}, no conclusive evidence exists.

The submillimeter band is almost the only wavelength range where obscuration is not an issue, motivating MBHB follow-up with a submillimeter instrument.
%
Gravitational wave observatories pinpoint sources' position within large regions of 10 deg$^2$ or more, far too wide for follow-up with interferometric instruments. 
The large FoV of AtLAST is key to improve the sky localization of gravitational wave detectors (LGWA, LISA); combined with its fast scanning speed, AtLAST can easily identify the source in the spiraling phase before the coalescence, when the MBHB system is expected to periodically varying in a few hundred/thousand seconds (depending on the masses). 

\section{Instrumental Requirements}
While the instrumental requirements are not the same between the assorted transient and variable science cases, there are a number of stringent specifications that are set forth by multiple cases. First, nearly all of the cases benefit from a large FoV. In the case of commensal observations, the larger FoV leads directly to a larger number of detected objects, as the rate of these serendipitous observations is naturally set by rate of sky coverage. Additionally, for targeted field observations, a larger FoV enables faster mapping, and hence less time required to meet a sensitivity target. ToO observations would benefit from a large FoV if the localization of the target is only poorly known (e.g., gravitational wave events: 10s to 100s of square degrees); in addition, a large FoV can allow for observation of a calibrating source while on target, reducing calibration overheads. 

Additionally many of the cases require simultaneous observations in several polarization sensitive bands. In some cases the simultaneous requirement is strict, such as for GRBs, while in other cases there is a large science gain but not a strict requirement, such as with asteroids. Many of these cases also require or greatly benefit from polarization sensitive observations. The GRB science case sets the most stringent requirement on the bands themselves, needing at least three bands spanning an order of magnitude in frequency (i.e. from $90$-$850$\,GHz). Other cases such as asteroids benefit from having as many bands a possible, but do not require any specific number.

The ToO science case is a particularly exciting one, with upcoming experiments like Vera Rubin promising a deluge of real-time alerts for rapid transients. Taking advantage of these alerts, however, requires that AtLAST be able to get on target in a matter of minutes. While this is mostly an operational requirement (see Section~\ref{sec:ded_obs}), AtLAST does need to be able to slew fast enough ($\sim 1$\,degree per second) to reach ToOs in a timely manner. More stringently, AtLAST would ideally not need to refocus after a fast slew to a ToO. This is an exacting demand on the mirror backing, and the trade off between slew speed and refocusing time needs to be studied.

\section{Observational Requirements}
In addition to instrumental requirements, the transient and variable case feeds back on AtLAST's operational and observation strategy. Both slowly varying and flaring sources, as well as triggering events, place unique constraints on observing cadence, duration, and sky coverage. Some of the science cases above justify their own, dedicated surveys. In other cases only commensal observations are needed. Maximizing the science yield of commensal observations, however, still requires some tuning of observational parameters and data processing procedures. In this section, we discuss both some possible dedicated surveys motivated by transient science, as well some considerations to take the maximum advantage of commensal observations.


\subsection{Dedicated Observations}
\label{sec:ded_obs}
Wide-field surveys with AtLAST can potentially revolutionize searches for transient millimeter-band sources.  If the volume density of a class of objects is constant over the possible ranges of survey depth, then the event rate will scale with the volume of the survey, which will be $\frac{\Omega}{3}\, d_{max}^3$ (ignoring relativistic effects), where $\Omega$ is the solid angle of the survey and $d_{max}$ is the maximum distance to which the phenomenon can be discovered. The exposure time needed will scale as $\Omega \, d_{max}^{4}$, so for a fixed exposure time, $\Omega$ will scale as $d_{max}^{-4}$, and the event rate will scale as $d_{max}^{-1}$. This highlights a preference for wider, shallower surveys over narrower, deeper ones for transient detection. 

This preference can be overridden for events which are not homogeneously distributed over the survey region. Some classes of objects will show strong evolution in number density with redshift (e.g. phenomena closely associated with star formation or AGN activity) or with distance from the Galactic Center (most Galactic phenomena).  Because of this, a range of surveys with different cadences, durations, solid angles and locations may be desirable.  In many cases, though, it will be possible to do such transient searches commensally in deep fields observed for other purposes by spreading the deep field observations out with an optimal cadence.

While AtLAST is capable of reaching very high sensitivity in mere seconds on a source, the transient science case specifically motivates longer exposure times. The atmosphere is highly variable at timescales of several seconds, motivating observing periods of $\geq 1$\,minute to average down those variations. Moreover, anomalous refraction leads to extrinsic time variation of point source flux, further requiring longer observations to average out such fluctuations. As such, for transient science, surveys with $\geq1$\, minute of observation time per position are preferred. 



{\bf Protostar Monitoring}
\label{sec:yso_survey}

As an example of the `capability of AtLAST, one potential wide-field survey for AtLAST could be a variable protostar monitoring survey patterned on the JCMT transient survey. In a nominal configuration, this survey would cover the Galactic plane, the Magellanic Clouds, and several out-of-plane star forming regions totaling approximately $550$ square degrees. Observing these fields at one square degree per minute would require approximately 15 hours and would yield three minutes of observing per pointing. This relatively short observing allocation would be appropriate for monthly monitoring of the fields. Using bandwidths tuned to provide the best sensitivity from a given atmospheric transmission window (as set out in  \cite{DiMascolo2024}), this would yield a sensitivity of $110\,\mu$Jy per beam at $350$\,GHz and $830\,\mu$Jy per beam at $650$\,GHz as compared to $12\,$mJy per beam and $500\,$mJy per beam for the JCMT Transient Survey. Such an increase in sensitivity would be transformative for understanding the accretion properties of young stars.

{\bf Targets of Opportunity}
AtLAST will be uniquely poised to provide submillimeter follow-up observations of transient events detected at other wavelengths. In general getting AtLAST on a source would take about an hour from an arbitrary observational configuration. However, when the wide-bandwidth continuum instrument is on sky, a rapid response mode (RRM) can be implemented. For some transient events, such as FBOTs \citep{bright2022} and GRBs \citep{bright2022}, the first few hours of observations are critical for characterizing the event. In RRM, response times of minutes can be achieved using automatic triggering by external alerts, for example from Vera Rubin or IceCube. A large FoV is critical for optimal RRM observations; with alarge FoV, small pointing errors can be corrected for post-observations. Moreover, calibration scans can be interleaved after the first on-source scan, with the potential to use the large FoV to opportunistically calibrate while scanning on source. The most pressing concern after a rapid slew would be focus; as such either methods need to be developed to maintain focus over a rapid slew, or a rapid focusing strategy needs to be developed. In addition to the technical requirements, procedures must be developed to rapidly evaluate the technical and scientific merit of the ToO. To some extent this can be externalized as, for example, the Rubin alert brokers will categorize the transient events detected, so that AtLAST will only need to determine ahead of time which events are worth of RRM within minutes as opposed to standard follow-up on the scale of hours.

In addition to automatically triggered observations in RRM mode, AtLAST should accept proposals for semi-automatic triggered observations. These proposals should have stringent requirements to allow AtLAST to most rapidly get on target. In this regard, the ALMA and GBT ToO guidelines are useful, for example requiring observing modes and target sensitivities to be specified prior to the proposal being triggered. The proposal could then be triggered manually by the PI, with observations following shortly thanks to pre-planned observations.

{\bf Coordinated observations}\\
Many of the classes of objects that exhibit dramatic, exciting transient source phenomenology are extremely broad spectrum sources, varying from radio through gamma-rays.  Developing deep physical understandings of these sources almost always requires multi-wavelength data, and sometimes requires coordinated observations. AtLAST will be uniquely suited to this task, being able to provide continuous, multi-chroic scheduled observations with significantly reduced need to go off-target to calibrate as compared to an interferometric instrument. AtLAST will need to have the capability to schedule both coordinated monitoring campaigns and strictly simultaneous long runs from its start. This will require agreements with other observational facilities, agreements which should be set before the beginning of operations, so that observational strategy can be decided ahead of time.





\subsection{Commensal Observations}
{\bf Ecliptic Plane}

As discussed above, observations of asteroids, comets, and TNOs can be completely opportunistic. Stacking of commensal observations of asteroids has already shown success \citep{Chichura2022, Orlowski-Scherer+23}, and similar approaches should work for both TNOs and comets. The only restriction is that observations near the ecliptic are more fruitful, as asteroids and TNOs generally occupy fairly low inclination orbits. Main belt asteroids spend $80\%$ of their time between a declination of $\pm 25\deg$. Within this declination range, the observing strategy is not particularly important. This is not an overly arduous restriction; for comparison, somewhat more than half of ACT observations fall between a declination of $\pm25\deg$ \citep{Naess2020}. Further, since there are a non-negligible number of asteroids and TNOs with highly inclined orbits, the transition to more extreme declination is relatively soft. One restriction of commensal observations is that the calibration, particularly on long time scales, be consistent, motivating regular observations of calibrators. 


In general, distinguishing intrinsic slow variability from extrinsic variability requires $\sim 1$\,minute of observation time (see Section~\ref{sec:ded_obs}). Ideally surveys would have array crossing times at least greater than this in order to maximize commensal slow variability science. In other words, surveys without a compelling reason to scan quickly should scan slowly. When this is not possible, while slowly varying transient science is not possible, commensal observations of rapidly varying sources such as GRBs is still possible, as are those of intrinsically invariant sources like asteroids and comets. 

{\bf Pipeline transient searches}\\
With a large FoV and very high instantaneous sensitivity, AtLAST holds the promise of blind detections of a large number of transient events. A such, a great benefit to the community would thus arise by developing a transient search pipeline to identify these transients in near real time, so that they can be reported almost immediately as a service to the community. Such pipelines are being developed for next generation CMB experiments such as SO  (Simons Observatory Collaboration, in prep.). At worst these alerts will be available on roughly daily timescales, but transient searches directly in the time ordered data hold the promise of $\sim$hour timescale alerting. For automated alerting, principal investigators should have an ``opt-out'' option if their core scientific goals are searches for transients in their fields. This would include both solar system objects and Galactic and extragalactic transients.

\subsection{Calibration}

For all monitoring programs, careful control of systematics will be necessary to achieve sufficient calibration accuracy. For detection of large amplitude bursts, standard submillimeter calibration techniques should suffice. However, for more subtle variability studies, such as protostar monitoring, it will be essential to attain state-of-the-art relative calibration levels reaching, or exceeding, 1\% uncertainty \citep{Mairs+2017a, Mairs+2024}. To maximally take advantage of commensal observations, calibration needs to be maintained over long periods of time. Specifically, regular observation of calibrating sources is required to disentangle intrinsic flux variation in a source from extrinsic calibration drift when commensal observations are separated by long periods of time, days or more. Regular planet scans will be critical for dialing in the absolute calibration. Additionally, the ALMA calibration targets are all visible from the AtLAST location, and have high density on the sky, at least at the lower frequencies, making them ideal for supplementing calibration by thermal emitters. Finally, it may be possible to reduce the short timescale fluctuations in calibration by simultaneously measuring the precipitable  water vapor (PWV) and using this measurement to remove short time-scale variations. ALMA currently uses a similar method to apply its atmospheric phase correction \citep{Nikolic2013}. As such, it is worthwhile to have a PWV radiometer integrated into AtLAST from the start, to enable this correction if possible.

\section{Summary of Instrumentation and Observational Requirements}
In order to gain a transformational improvement in our ability to study the transient submillimeter sky, a single-dish telescope is needed with a large FoV, high sensitivity, high angular resolution, multi-chroic camera, and an ability to promptly respond to targets of opportunity. A large FoV will enable commensal observations of transient events during scientifically unrelated surveys. Moreover, a large FoV enables a high scanning speed, which permits high cadence surveying of interesting transient fields, such as star forming regions. High sensitivity and angular resolution combined are required to detect the faintest transient events, which are generally point sources and hence subject to beam dilution. The ability to make simultaneous, multi-chroic observations is critical for several science cases such as GRBs and FBOTs. Finally, a single dish instrument can promptly respond to targets of opportunity, obtaining data within an hour of a triggered observation. All of these capabilities can be fulfilled by AtLAST \citep{Mroczkowski2024}. 

In addition, obtaining the best transient science results will require consideration in observing strategy. For some science cases, dedicated surveys will need to be planed, perhaps as subsets of larger surveys. For ToO science, a prompt response and calibration plan will need to be developed, which will allow proposers to quickly trigger observations of time-sensitive transient events. To maximize AtLAST's ability to follow-up transient events (i.e. response times $\lesssim1$\,hour), a fully automated pipeline triggering observations should be developed. For commensal observations, regular calibration will need to be implemented in order to separate intrinsic variability in detected sources from extrinsic calibration drifts. Finally, an alerting system will need to be put into place, which will alert members of the astronomical community to transients detected by AtLAST. The reward for this work will be a transformational improvement in our understanding of the transient submillimeter sky.

\section*{Data and software availability}

The calculations used to derive integration times for this paper were done using the AtLAST sensitivity calculator, a deliverable of Horizon 2020 research project `Towards AtLAST', and available from \href{https://github.com/ukatc/AtLAST_sensitivity_calculator}{this link}.

\section*{Grant information}
This project has received funding from the European Union’s Horizon 2020 research and innovation program under grant agreement No.\ 951815 (AtLAST).

S.W. acknowledges support by the Research Council of Norway through the EMISSA project (project number 286853) and the Centres of Excellence scheme, project number 262622 (``Rosseland Centre for Solar Physics''). 

F.M.M. acknowledges support by UCM through the María Zambrano grant, funded by the Spanish government with Next Generation EU funds. Grants PID2022-138621NB-I00 and RTI2018-096188-B-I00 are funded by Spanish MICIU and by the European Union.

C.R. acknowledges support from Fondecyt Regular grant 1230345 and ANID BASAL project FB210003.

P.S., C.V. and C.C. acknowledge a financial contribution from
the Bando Ricerca Fondamentale INAF 2022 Large Grant, ‘Dual and binary supermassive black
holes in the multi-messenger era: from galaxy mergers to gravitational waves’.

This work was supported by a grant from the Simons Foundation (CCA 918271, PBL).

\section*{Acknowledgements}
We would like to acknowledge the numerous people who discussed with us the potentialities for the submillimeter transient sky.


\begingroup
\small
\bibliographystyle{apj_mod}
\setlength{\parskip}{0pt}
\setlength{\bibsep}{0pt}
\bibliography{atlast}
\endgroup


\end{document}